\documentclass[twocolumn,twoside,slac_two]{revtex4}
\usepackage{graphicx}
\usepackage{fancyhdr}
\begin{document}

\title{On the Magnetic Field in Kpc-Scale Jets of FR I Radio Galaxies}

\author{\L . Stawarz, A. Siemiginowska}
\affiliation{Harvard-Smithsonian Center for Astrophysics, 60 Garden Street, Cambridge, MA 02138, USA}

\author{M. Ostrowski}
\affiliation{Astronomical Observatory, Jagiellonian University, ul. Orla 171, 30-244 Krak\'ow, Poland}

\author{M. Sikora}
\affiliation{Nicolaus Copernicus Astronomical Center, ul. Bartycka 18, 00-716 Warszawa, Poland}

\begin{abstract}
The energy content of large-scale jets in FR I radio galaxies is still an open issue. Here we show that upper limits on the high-energy and very high-energy $\gamma$-ray emission of the kpc-scale jet in M~87 radio galaxy imposed by {\it EGRET}, {\it Whipple}, and --- most importantly --- {\it HEGRA} and {\it HESS} observations, provide important constraints on the magnetic field strength in this object. In particular, a non-detection of $\gamma$-ray radiation from the brightest part of this jet (knot A), expected from the inverse-Compton scattering of the starlight photons by the synchrotron-emitting jet electrons, implies that the magnetic field cannot be smaller than the equipartition value (referring solely to the radiating ultrarelativistic electrons), and most likely, is even stronger. In this context, we point out several consequences of the obtained result for the large-scale jet structures in FR I radio galaxies and the M~87 jet in particular. For example, we discuss a potential need for amplification of the magnetic field energy flux along these jets (from sub-pc to kpc scales), suggesting a turbulent dynamo as a plausible process responsible for the aforementioned amplification. 
\end{abstract}

\maketitle

\thispagestyle{fancy}

\section{INTRODUCTION}

Stawarz et al. (2003) considered very high energy (VHE) $\gamma$-ray emission produced by the kpc-scale jets in nearby low-power radio galaxies of the FR I type~\cite{sta03}. Optical and X-ray emission detected recently from a number of such objects indicate that these jets are still relativistic on the kpc scale, and that they contain ultrarelativistic electrons with energies up to $100$ TeV (see a discussion in~\cite{sta04} and references therein). Therefore, some of the nearby FR I jets can be in principle VHE $\gamma$-ray emitters due to the inverse-Compton (IC) scattering off ambient photon fields which, at kpc distances from the active nuclei, are expected to be still relatively high. For example, following~\cite{tsa95} it can be found that the bolometric energy density of the stellar emission at $1$ kpc from the center of a typical luminous elliptical galaxy is on average $U_{\rm star, \, bol} \approx 10^{-9}$ erg cm$^{-3}$. In the particular case of radio galaxy M~87,~\cite{sta03} show that comptonization of such a starlight radiation (that dominates over the energy densities of the other photon fields in the jet comoving frame, in particular over the energy density of the synchrotron photons) within the kpc-scale jet (its brightest knot A) by the synchrotron-emitting electrons in the equipartition magnetic field can possibly account for the TeV emission detected by the {\it HEGRA} Cherenkov Telescope from the direction of that source~\cite{aha03}. However, subsequent observations of M~87 by the {\it Whipple} Telescope~\cite{leb04} gave only upper limits for its emission in the $0.4 - 4$ TeV photon energy range, suggesting, although not strictly implying, variability of the VHE $\gamma$-ray signal. Such variability, clearly confirmed by the most recent {\it HESS} observations --- which established the presence of a variable source of VHE $\gamma$-ray emission within 0.1 deg ($\sim 30$ kpc) of the M~87 central region~\cite{bei05} --- questions the possibility that the kpc-scale jet is responsible for the $3 - 4 \sigma$ detections of M~87 by {\it HEGRA} and {\it HESS} telescopes. On the other hand, the upper limits imposed in this way put interesting constraints on the magnetic field within the M~87 large-scale jet, an issue which is of general importance in astrophysics.

Here we comment on the high energy $\gamma$-ray emission of the M~87 kpc-scale jet, resulting from the IC scattering on the stellar photon field. We take into account a relativistic bulk velocity of the emitting region as well as Klein-Nishina regime of the electron-photon interaction. We emphasize an important aspect of the presented discussion: IC scattering of the starlight emission by the synchrotron-emitting electrons is \emph{inevitable}, and involves neither the unknown target photon field, nor the additional unconstrained source of the ultrarelativistic high-energy electrons. In particular, we `reconstruct' the electron energy distribution from the \emph{known} broad-band synchrotron spectrum of a given jet region, and then estimate the IC flux for the \emph{known} target photon field. Therefore, our results, presented in detail in~\cite{sta05} are independent on any model of particle acceleration.

\section{THE MODEL}

\subsection{Electron energy distribution}

Detailed broad-band (radio-to-X-ray) observations of the M~87 large-scale jet circumscribe well the synchrotron spectrum of knot A, the brightest part of the jet in the radio and optical regimes. Hereafter we refer to the analysis by~\cite{wil02}, which indicates that the energy distribution of the synchrotron-emitting electrons (as measured in the jet commoving frame denoted by primes) can be approximated by a broken power-law,
\begin{equation}
n'_{\rm e}(\gamma) \propto \left\{\begin{array}{ccc} \gamma^{-p} & {\rm for} & \gamma \leq \gamma_{\rm br} \\ \gamma_{\rm br}^{q} \, \gamma^{-(p + q)} & {\rm for} & \gamma > \gamma_{\rm br} \end{array} \right. \quad ,
\end{equation}
\noindent
with the spectral indices $p = 2.3$ and $q = 1.6$. The electron break Lorentz factor $\gamma_{\rm br} \approx 2.7 \times 10^6 \, \delta^{-0.5} \, B_{-4}^{-0.5}$ corresponds to the observed synchrotron break frequency $\nu_{\rm br} \approx 10^{15}$ Hz for the emitting plasma magnetic field $B \equiv B_{-4} \, 10^{-4}$G and Doppler factor $\delta$. The cut-off energies $\gamma_{\rm min}$ and $\gamma_{\rm max}$ are basically unconstrained, but the synchrotron origin of the X-ray jet emission indicates $\gamma_{\rm max} / \gamma_{\rm br} \geq 10 - 100$. We normalize the number of radiating electrons to the observed (isotropic) synchrotron luminosity of knot A at the break frequency, $L_{\rm br} \approx 3 \times 10^{41}$ erg s$^{-1}$~\cite{wil02}.

\subsection{Starlight photon field}

For the target starlight photons at the position of knot A we assume roughly isotropic distribution in the galactic rest frame and strongly anisotropic in the jet rest frame, due to the relativistic jet velocity. We take the characteristic observed frequency of the optical-NIR bump due to the elliptical host of M~87 as $\nu_{\rm star} \approx 10^{14}$ Hz~\cite{mul04}. We also evaluate the appropriate starlight energy density at the position of knot A \emph{directly} from the observations of M~87 host galaxy, assuming King profile for the stellar emissivity at the distances $>1'' - 2''$ from the M~87 nucleus~\cite{you78},
\begin{equation}
j_{\rm star}(r) \propto \left[ 1 + \left({r \over r_{\rm c}}\right)^2\right]^{-3/2} \quad {\rm for} \quad r < r_{\rm t} \quad ,
\end{equation}
\noindent
where $r$ is the radius as measured from the galactic center, $r_{\rm c}$ is the core radius for the galaxy and $r_{\rm t}$ is the appropriate tidal radius. We normalize this distribution to the observed luminosity density profile in the $I$ band~\cite{lau92}, with the parameters $r_{\rm c} = 0.55$ kpc and $r_{\rm t} = 68$ kpc. We evaluate further the $I$-band stellar energy density at the position of knot A as $U_{\rm star}(1 \, {\rm kpc}) \approx 10^{-10}$ erg cm$^{-3}$, which is in fact a very safe \emph{lower} limit (see a discussion in~\cite{sta05}).

\subsection{IC emission and kinematic factors}

We evaluate the high-energy emission of knot A due to IC scattering on monoenergetic and mono-directional (in the jet rest frame) starlight photon field, including properly relativistic effects in the Klein-Nishina regime, using the approximate expression given by~\cite{aha81}. We also consider the bulk Lorentz factor of the M~87 jet at the position of knot A, $\Gamma$, to be $\sim 3 - 5$, and the jet viewing angle $\theta$, $\sim 20^0 - 30^0$~\cite{bic96}. For such a choice of $\Gamma$ and $\theta$ one gets the jet Doppler factor of the discussed region $\delta = [\Gamma (1 - \beta \cos \theta)]^{-1} \sim 1 - 3$.

\section{THE RESULTS}

Spectral energy distribution of the high-energy $\gamma$-ray IC emission of knot A is presented in figure 1 for different magnetic field intensities, jet viewing angles $\theta = 30^0$, $20^0$, and the bulk Lorentz factors $\Gamma = 3$ and $5$. The Thomson part of this emission extends up to photon energies of about $10^{10} - 10^{11}$ eV. Below we compare the expected IC fluxes for different photon energies to the upper limits imposed by the observations of {\it EGRET} observatory and ground-based Cherenkov Telescopes: {\it HESS}, {\it HEGRA} and {\it Whipple}, and next we compare the implied lower limits on the jet (knot A) magnetic field with the appropriate equipartition value, 
\begin{equation}
B_{\rm eq} = 330 \, \delta^{- 5/7} \, \mu{\rm G} \quad ,
\end{equation}
\noindent
as given by~\cite{kat05}. Note, that the adopted value of $B_{\rm eq}$ refers to the energy equipartition between the jet magnetic field and ultrarelativistic electrons, with possible contribution from the non-radiating particles neglected. 

\begin{figure*}[t]
\centering
\includegraphics[width=135mm]{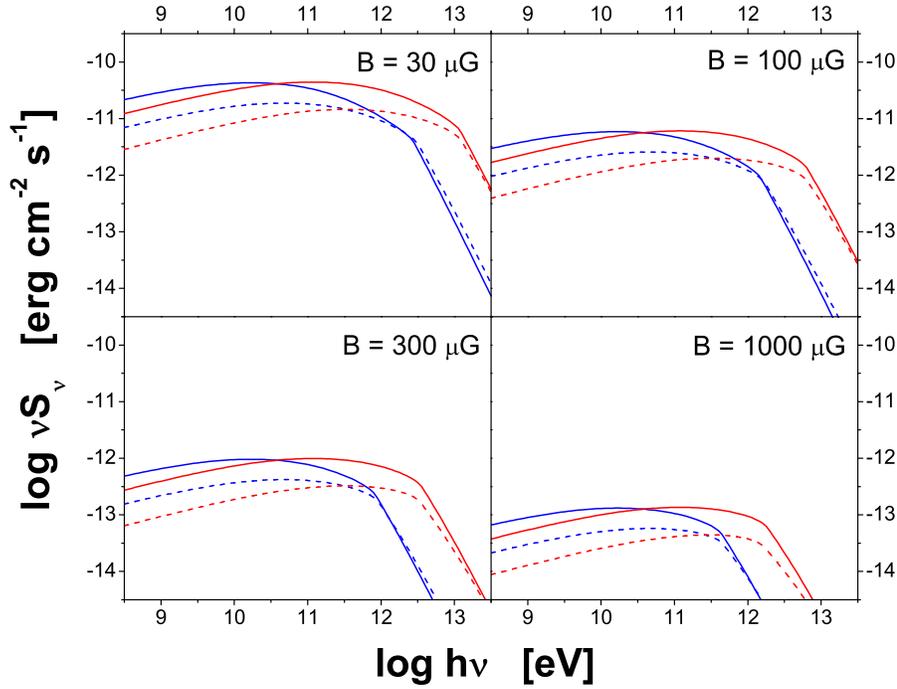}
\caption{Spectral energy distribution of high-energy $\gamma$-ray inverse-Compton emission of knot A for the jet magnetic field $B = 0.3 \times 10^{-4}$ G, $10^{-4}$ G, $3 \times 10^{-4}$ G, $B = 10^{-3}$ G (as indicated at each panel), the jet viewing angles $\theta = 30^0$ and $20^0$ (blue and red curves, respectively) and the jet bulk Lorenz factors $\Gamma = 5$ and $3$ (solid and dashed lines, respectively).}
\end{figure*}

\subsection{{\it EGRET} observations}

{\it EGRET} observations of the Virgo cluster give the photon flux $F(> 100 \, {\rm MeV}) < 2.2 \times 10^{-8}$ cm$^{-2}$ s$^{-1}$~\cite{rei03}, which implies $B > 30 - 100$ $\mu$G corresponding roughly to $B / B_{\rm eq} > 0.1 - 0.5$, for any choice of the kinematic factors considered here. This constraint does not necessarily imply a departure from the magnetic field--radiating particles energy equipartition but, interestingly enough, already excludes a class of models involving a very weak jet magnetic field.
 
\subsection{{\it Whipple} observations}

{\it Whipple} observations give the $99 \%$ CL upper limit to the VHE $\gamma$-ray photon flux of M~87 radio galaxy $F(> 0.4 \, {\rm TeV}) < 6.9 \times 10^{-12}$ cm$^{-2}$ s$^{-1}$~\cite{leb04}. Independently on the {\it EGRET} constraints, the {\it Whipple} observations imply thus magnetic field intensity within the discussed jet region $B > 50 - 60$ $\mu$G, or in other words $B / B_{\rm eq} > 0.2 - 0.4$ again for any choice of the kinematic factors considered here.

\subsection{{\it HEGRA} and {\it HESS} observations}

{\it HEGRA} observations resulted in a $4 \sigma$ detection of high-energy $\gamma$-ray flux from the direction of M~87 with the photon flux $F(> 0.73 \, {\rm TeV}) \approx 0.96 \times 10^{-12}$ cm$^{-2}$ s$^{-1}$~\cite{aha03}. Recent {\it HESS} observations resulted also in the detection of the M~87 system at the $3 - 4 \sigma$ level, however with the photon fluxes $F(> 0.73 \, {\rm TeV}) \approx 0.4 \times 10^{-12}$ cm$^{-2}$ s$^{-1}$ in 2003, and $F(> 0.73 \, {\rm TeV}) \approx 0.15 \times 10^{-12}$ cm$^{-2}$ s$^{-1}$ in 2004~\cite{bei05}. This clearly indicates a variability of the high-energy $\gamma$-ray emission of this source, and therefore gives the \emph{upper limits} for the VHE radiation of knot A.

\begin{figure*}[t]
\centering
\includegraphics[width=135mm]{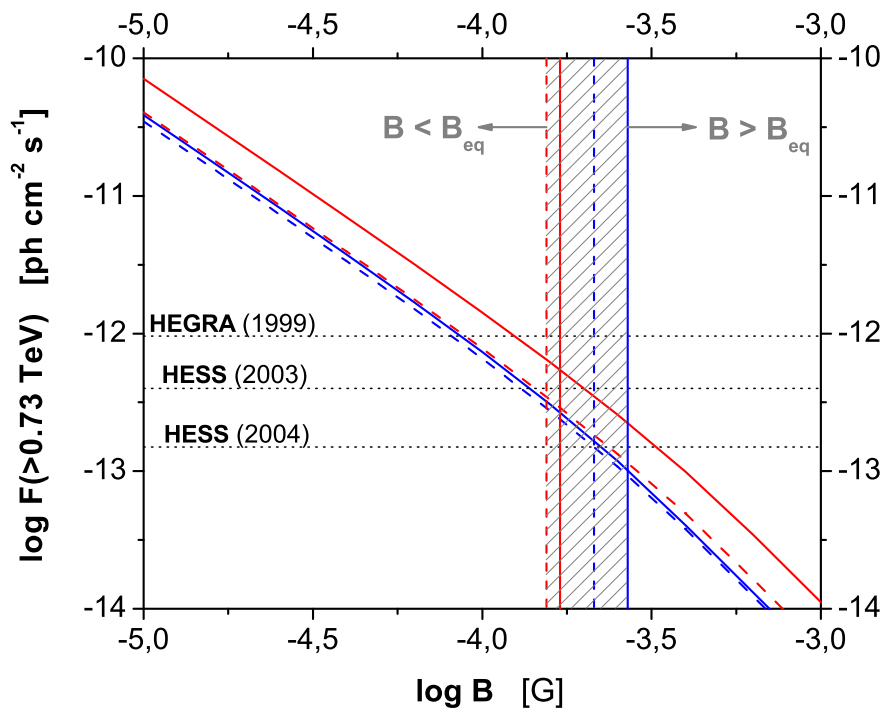}
\caption{Constraints on the jet magnetic field within knot A imposed by the {\it HEGRA} and {\it HESS} observations (dotted horizontal lines), for the jet viewing angle $\theta = 30^0$ and $20^0$ (blue and red curves, respectively), and the jet bulk Lorenz factors $\Gamma = 5$ and $3$ (solid and dashed lines, respectively). Vertical lines denote the equipartition magnetic field for $\Gamma = 5$ and $3$ (solid and dashed lines, respectively) and $\theta = 30^0$ and $20^0$ (blue and red curves, respectively).}
\end{figure*}

Figure 2 shows the expected photon flux of knot A at the observed photon energies $ h \nu_0 > 0.73$ TeV as a function of the magnetic field intensity, for the jet viewing angles $\theta = 30^0$ and $20^0$, and the bulk Lorentz factors $\Gamma = 3$ and $5$. The vertical lines indicate the appropriate equipartition magnetic field. The most recent {\it HESS} observations imply therefore the magnetic field within knot A as strong as $B > 300$ $\mu$G, i.e. $B / B_{\rm eq} > 1 - 2$. 

\section{CONCLUSIONS}

Our study indicates that the magnetic field within the brightest knot A of M~87 jet is most likely stronger than the equipartition value that refers to the radiating electrons alone, $B \gtrsim 300 \, \mu{\rm G} \gtrsim B_{\rm eq}$ (if the jet viewing angle is in the range $\theta = 20^0 - 30^0$ and the jet bulk Lorentz factors $\Gamma = 3 - 5$). On the other hand the upper limit of the magnetic field intensity within knot A can be found from the upper limit imposed on the magnetic field energy flux, $L_{\rm B} \equiv {1 \over 8} R^2 c \Gamma^2 B^2 \leq L_{\rm j}$, where $L_{\rm j} \sim {\rm few} \times 10^{44}$ erg s$^{-1}$ is the total power of M~87 jet~\cite{owe00} and $R \approx 60$ pc is the radius of radio structure at the position of knot A. This gives roughly $B_{\rm max} < 1000$ $\mu$G. The comoving energy density of the magnetic field within knot A is therefore limited roughly by $4 \times 10^{-8}$ erg cm$^{-3}$ $\geq U'_{\rm B}$ $\geq 4 \times 10^{-9}$ erg cm$^{-3}$.

{\it VLBI} measurements often allow one to infer magnetic field intensity from the low-frequency spectral break in the radio emission of the (sub) pc-scale jet modeled in terms of synchrotron self-absorption process. In the case of M~87 jet, this method gives $B_{\rm VLBI} < 0.2$ G at $R_{\rm VLBI} \ll 0.06$ pc~\cite{rey96}. If the magnetic energy flux in a jet is constant, then one should expect magnetic field intensity at the position of knot A to be roughly $B = (\Gamma_{\rm VLBI}/ \Gamma) \, (R_{\rm VLBI} / R) \, B_{\rm VLBI}\ll 300$ $\mu$G, where we put $\Gamma_{\rm VLBI} / \Gamma \approx 2$ and the jet radius at the position of the considered knot $R = 62$ pc~\cite{owe00}. Hence one can conclude that some kind of amplification of the jet magnetic field has to take place between the parsec and kiloparsec scales, although all the estimates presented above should be taken with caution, as some arbitrary assumptions on the jet magnetic field structure were invoked.

We speculate that the suggested amplification of the jet magnetic field is due to an action of the turbulent dynamo processes discussed in this context by, e.g.,~\cite{dey80}: the Kelvin-Helmholtz instabilities occurring inevitably at the edges of the jet are supposed to create large-scale eddies at the flow boundaries, which amplify the jet magnetic field and develop highly turbulent mixing layer between the jet and the surrounding medium. Such turbulent shear layers play a crucial role in mass entrainment (and so deceleration) of the FR I outflows~\cite{bic94}, and in acceleration of the jet particles~\cite{sta02}, influencing also polarization properties of the jets~\cite{lai80}. 

In a framework of this model, on the small scales the M~87 jet has to be dynamically dominated by the (cold) particles, in order to allow for the turbulent dynamo process to proceed at all. On the larger scales the mass entrainment process decelerates the flow gradually, slowly amplifying the jet magnetic field (to the value exceeding at some point the energy equipartition with the radiating electrons), and creating a turbulent boundary layer that spreads into the jet interior. This process accomplishes at $\sim 1 - 2$ kpc from the active center (knot A and beyond), where the jet magnetic field reaches an approximate equipartition with the energy density of the jet particles, while the jet flow itself becomes fully turbulent, disappearing into the outer amorphous radio lobe.

\bigskip
\begin{acknowledgments}
\L .S. was supported by the grant 1-P03D-011-26 and partly by the Chandra grants G02-3148A and G0-09280.01-A. A.S. was supported by the NASA contract NAS8-39073 and Chandra Avard G02-3148A. M.O. and M.S. were supported by the grant PBZ-KBN-054/P03/2001. \L .S. acknowledges also very useful comments from F. Aharonian, C.C. Cheung, D.S. De Young, D.E. Harris, and J. Kataoka. 
\end{acknowledgments}

\end{document}